# All-optical sampling and magnification based on XPM-induced focusing

J. Nuño[1,a)] M. Gilles[1], M. Guasoni[1-2], C. Finot[1], and J. Fatome[1]

[1] *Laboratoire Interdisciplinaire Carnot de Bourgogne (ICB), UMR 6303 CNRS-Univ. Bourgogne Franche-Comté, 9 av. A. Savary, 21078 Dijon, France.*

[2] *Nonlinear Physics Center, Australian National University, Canberra ACT 0200, Australia.*

We theoretically and experimentally investigate the design of an all-optical noiseless magnification and sampling function free from any active gain medium and associated high-power continuous wave pump source. The proposed technique is based on the co-propagation of an arbitrary shaped signal together with an orthogonally polarized intense fast sinusoidal beating within a normally dispersive optical fiber. Basically, the strong nonlinear phase shift induced by the sinusoidal pump beam on the orthogonal weak signal through cross-phase modulation turns the defocusing regime into localized temporal focusing effects. This periodic focusing is then responsible for the generation of a high-repetition-rate temporal comb upon the incident signal whose amplitude is directly proportional to its initial shape. This internal redistribution of energy leads to a simultaneous sampling and magnification of the signal intensity profile. This process allows us to experimentally demonstrate a 40-GHz sampling operation as well as an 8-dB magnification of an arbitrary shaped nanosecond signal around 1550 nm in a 5-km long normally dispersive fiber. The experimental observations are in quantitative agreement with numerical and theoretical analysis.

In modern photonic systems, the sampling process has widespread applications in the fields of optical communications, metrology, clocking, sensing, spectral comb or arbitrary waveform generation. In this context, nonlinear effects have been demonstrated as potential key technologies to develop all-optical sampling devices[1-14]. In most of these techniques, an ultrashort pulse train acts as an optical gate and the basic physical phenomena under use include four-wave mixing (FWM)[10], cross-phase modulation (XPM)[11], nonlinear polarization rotation[13] and Raman soliton self-frequency shifting[14]. On the other hand, signal amplification is also a critical function in many area of physics. Basically, optical amplification process refers to multiply an incident signal by increasing its global energy through an active gain medium pumped by an external pump beam. The large range of current and mature amplification techniques at telecommunication wavelengths involves Erbium doped fiber amplifiers, Raman-based amplifiers,

_________________________

a) Electronic mail: j.nugno@gmail.com

semiconductor amplifiers or parametric processes. However, it is noteworthy that most of these techniques lead to an inherent degradation of the signal-to-noise ratio induced by the detrimental spontaneous emission of photons. More recently, Azaña and coworkers have suggested and experimentally validated a different scenario in which a linear redistribution of energy into a periodic pulse train through the self-imaging Talbot effect can lead to a noiseless amplification process[15]. In this manuscript, we propose an alternative nonlinear approach enabling to simultaneously sample and magnify an arbitrary shaped incident signal due to a focusing effect occurring in a normally dispersive medium through the cross-phase modulation induced by an orthogonally polarized high-repetition-rate sinusoidal pump beam. Fundamentally, the strong sinusoidal beating is responsible for a localized and periodic focusing of the signal energy into high repetition pulses generated at the pump modulation frequency[16]. This internal redistribution of energy within the signal wave leads to the generation of a temporal intensity comb whose amplitude is then directly proportional to the signal profile, thus inducing a simultaneous sampling and magnifying process. Based on this technique, we have experimentally demonstrated a 40-GHz sampling operation as well as an 8-dB magnification factor of an arbitrary shaped nanosecond signal around 1550 nm in a 5-km long normally dispersive fiber. We have also derived a theoretical estimation of the total magnifying factor of the system in good agreement with our experimental observations and numerical simulations based on a simplified Manakov modelling.

Basically, it is well-known that the anomalous dispersive regime of propagation in optical fibers can lead to a focusing effect owing to the interplay between chromatic dispersion and Kerr nonlinearity. This process has been exploited in numerous studies in order for instance to compress optical pulses or design high-repetition pulse sources[17-19]. More recently, we have reported the observation of a focusing effect occurring in the normal dispersion regime through a XPM interaction between a weak periodic signal and its orthogonally polarized interleaved high-power replica[16]. In this previous configuration, a 3.3-ps 40-GHz pulse train was then generated[16]. In this new contribution, we exploit a similar physical mechanism[20] to induce a temporal focusing of an arbitrary shaped signal so as to sample and magnify its intensity profile. As a rule of thumb, the signal to be processed is co-propagating in a normally dispersive optical fiber together with a sinusoidal pump beam. In the present configuration, the pump wave consists in a much more intense orthogonally polarized high-repetition-rate carrier-suppressed beat-signal. Due to the nonlinear defocusing dynamics of the fiber, the pump is reshaped into parabolic then broad and sharp square pulses. Subsequently, optical shocks associated to wave-breaking phenomenon appear in its intensity profile inducing deep singularities at its null points[21]. As these singularities propagate, they are characterized by steeper and steeper edges, and then progressively collapse. In parallel, the XPM nonlinear phase shift induced by these nonlinear dark structures combined with the group velocity dispersion of the fiber then turns out the normal dispersion regime into a periodic focusing dynamics for the co-propagating signal. As a consequence, a fast and periodic pulse train exactly proportional to the incident signal is then generated, thus inducing a discrete sampling and an overall magnification effect of its intensity profile.

Figures 1 (a-c) show a numerical illustration of this process. For these simulations, we consider two orthogonally polarized optical waves propagating in a normally dispersive randomly birefringent fiber. The evolution of the complex slowly varying amplitudes of the signal $u$ and pump $v$ are described by the following set of two coupled nonlinear Schrödinger equations, corresponding to a simplified Manakov model in which nonlinear terms involved with the weaker signal have been neglected:

$$i\frac{\partial u}{\partial z} + \frac{\beta_2}{2}\frac{\partial^2 u}{\partial t^2} + \frac{8}{9}\gamma |v|^2 u + i\frac{\alpha}{2}u = 0,$$

$$i\frac{\partial v}{\partial z} + \frac{\beta_2}{2}\frac{\partial^2 v}{\partial t^2} + \frac{8}{9}\gamma |v|^2 v + i\frac{\alpha}{2}v = 0.$$

(1)

Here, $z$ and $t$ denote the propagation distance and time coordinates, respectively, whereas $\alpha$ indicate propagation losses. The factor 8/9 takes into account for random fluctuations of the intrinsic birefringence along the fiber length. In addition, we assume that third-order dispersion, losses or Raman effects play a minor role in the nonlinear dynamics and can therefore be neglected. We also assume that the wavelengths of the pump and signal under investigation are relatively close so that the walk-off between the two waves can be neglected. An arbitrary shaped weak signal is first injected into the fiber with a moderate average power (red solid line in Fig. 1b). The fiber parameters correspond to the fiber segment used in our experiment and is characterized by a chromatic dispersion $\beta_2 = 3.2$ ps$^2$/km and a nonlinear Kerr coefficient $\gamma = 1.7$ W$^{-1}$km$^{-1}$. A high power sinusoidal pump wave (26.5 dBm average power) is then simultaneously injected with an orthogonal polarization state (red line in Fig. 1a). Note that to ensure an efficient signal processing, the temporal period of the pump wave has to be much shorter than the signal duration. Owing to the strong nonlinear defocusing regime, the carrier-suppressed pump wave is drastically reshaped into the fiber and sharp singularities gradually appear at its null points (see blue line in Fig. 1a after 5 km of propagation). Simultaneously, we clearly observe in Fig. 1b that the focusing effect imposed by the co-propagating pump wave progressively leads to a redistribution of the signal energy. Indeed, at the output of the 5-km long fiber, an intensity comb at the pump repetition-rate is then generated upon the signal wave whose amplitude is directly proportional to its initial envelope, thus leading to both a discrete sampling and magnification effect (blue line in Fig. 1b). Finally, normalized profiles shown in Fig. 1c confirm that the profile of the magnified wave is a perfectly scaled version of the signal under test.

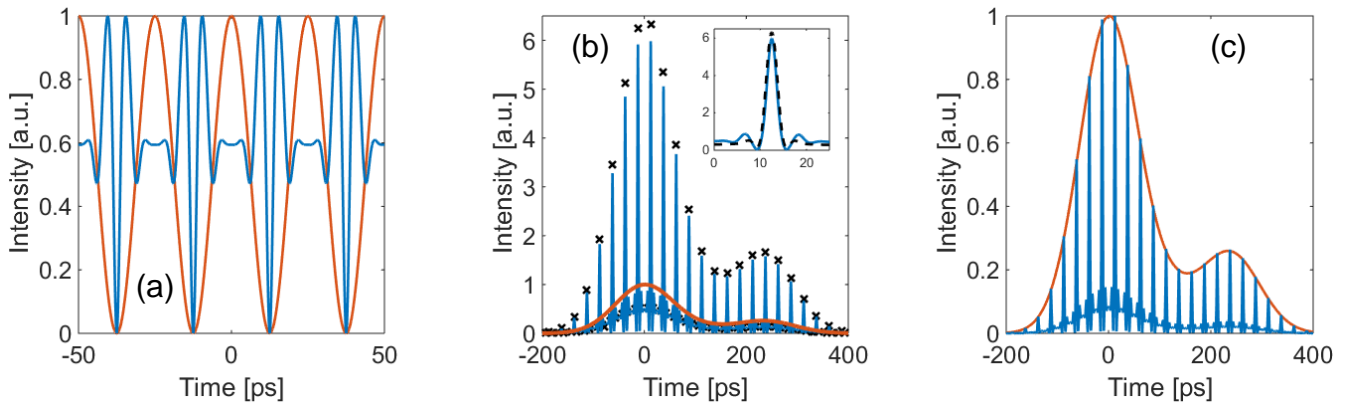

FIG. 1. Numerical simulations (a) Input (red) and output (blue) intensity profiles of the sinusoidal pump wave normalized to unity for an input average power of 26.5 dBm and 5 km of propagation. (b) Output intensity profiles of the orthogonally polarized signal wave when the pump is deactivated (in red) or activated (blue). The black crosses indicate the maximum values of the theoretical prediction. In the inset, a close view of the numerical and the theoretical estimations. (c) Same as (b) but normalized to unity.

In order to predict the magnification factor experienced upon the sampling process, it proves useful to develop a simplified model based on the analysis of the pump-signal interaction carried out in the optical spectral domain. The cosines-like input $v(0,t) \equiv v(t)_{in}$ pump reads as $v(t)_{in}=(P_0)^{1/2}cos(\omega_p t)$, to which correspond two spectral components centered at $\pm \omega_p$. The weak input signal is assumed to be a slowly-varying waveform $u(0,t) \equiv u_{in}(t)$, instead, whose corresponding optical spectrum is centered around $\omega=0$ with a spectral extension well below $\omega_p$. The FWM interaction among the pump and signal leads to the generation of new spectral components centered around odd (even) multiples of $\omega_p$ in the pump (signal) spectrum. On the other hand, the different power-level of pump and signal makes their longitudinal evolution deeply different. As long as the fiber length does not exceed few nonlinear lengths (defined as $1/\gamma P_0$), the level of the new spectral components generated by FWM in the pump spectrum remain 10 dB below the original components centered at $\pm \omega_p$[16]. We can thus neglect as a first rough approximation the impact of these additional spectral components and therefore consider that the sinusoidal pump intensity profile is preserved upon propagation except for propagation losses, that is $|v(z,t)|^2=P_0 \, cos(\omega_p t)^2 exp(-\alpha z)$. Note that this approximation remains still valid as long as the shock point undergone by the pump wave only occurs in the last stage of propagation, which is the case for most of the power values considered in the present experiment. In contrast, the generation of new spectral components cannot be neglected in the case of the weak signal, instead, as their amplitude rapidly grow up to the peak-level of the input signal spectrum $U(\omega)_{in}$. Replica of $U(\omega)_{in}$ arise in the signal spectrum, centered around even-multiples of $\omega_p$, to which the following ansatz $u(z,t)$ corresponds:

$$u(z,t) = u_{in}(t) \left[\sum S_{2n}(z) \, exp(i \, 2n \, \omega_p t)\right], \quad (2)$$

Insertion of the aforementioned ansatz in Eq.1 and subsequent collection of terms at the same frequency brings to the following linear system of differential equations (LDE): $\partial S_{2n}(z)/\partial z = (ia_{2n} - \alpha/2)S_{2n} + b(S_{2(n+1)} + S_{2(n-1)})$, being $a_{2n}=4n^2\omega_p^2\beta_2/2+(8/9)\gamma P$ and $b=(4/9) \, \gamma P$, where $P = P_0/L \int_{z=0}^{L} exp(-\alpha z)dz$ denotes the average pump power along the fiber length. In doing this, we have exploited the slowly-varying hypothesis for $u_{in}(t)$, which allows neglecting all terms in $\partial u_{in}(z)/\partial t$ and $\partial^2 u_{in}(z)/\partial^2 t$.

Solution of the LDE above requires truncation over a finite number of terms, that is $-N \leq n \leq N$, and the assignment of the input values $S_{2n}(0)$. In practice, N=4 (9 components) is sufficient to get a good estimation of $S_{2n}(z)$. Moreover the input spectrum $U(\omega)_{in}$ is centered around $\omega=0$, therefore we set $S_0(0)=1$ and $S_{2n}(0)=0$ for $n \neq 0$. Note that in the case of a symmetric spectrum, as in the present study, the modeling can be reduced to half of the truncated terms; that is to say only the $n \geq 0$ spectral components can be considered. It is interesting to notice that according to Eq. 2 the input beam $u_{in}(t)$ is modulated by a function $g(z,t) = \sum S_{2n}(z) exp(i2n\omega_p t)$. Solution of the LDE show that $g(z,t)$ evolves from a constant-in-time function $g(0,t)=1$ towards a periodic train of pulses that are progressively compressed along the fiber length. Consequently, the signal $u(z,t)$ experiences a simultaneous sampling and magnification, for which samples are more and more temporally compressed and intense upon propagation. In Fig. 1b, the maxima of the output intensity profile calculated by means of this analytical model have been reported with black crosses and compared to numerical simulations. We can notice a good agreement between both results, thus validating our theoretical approach. Furthermore, in the inset of Fig. 1b, a close view of both intensity profiles has been depicted and also reveals an excellent prediction of the generated pulse shape. Besides, as we will show in next section, this simplified analytic model well captures the features observed in experiments, and may allow optimizing and engineering the sampling and magnification processes here discussed.

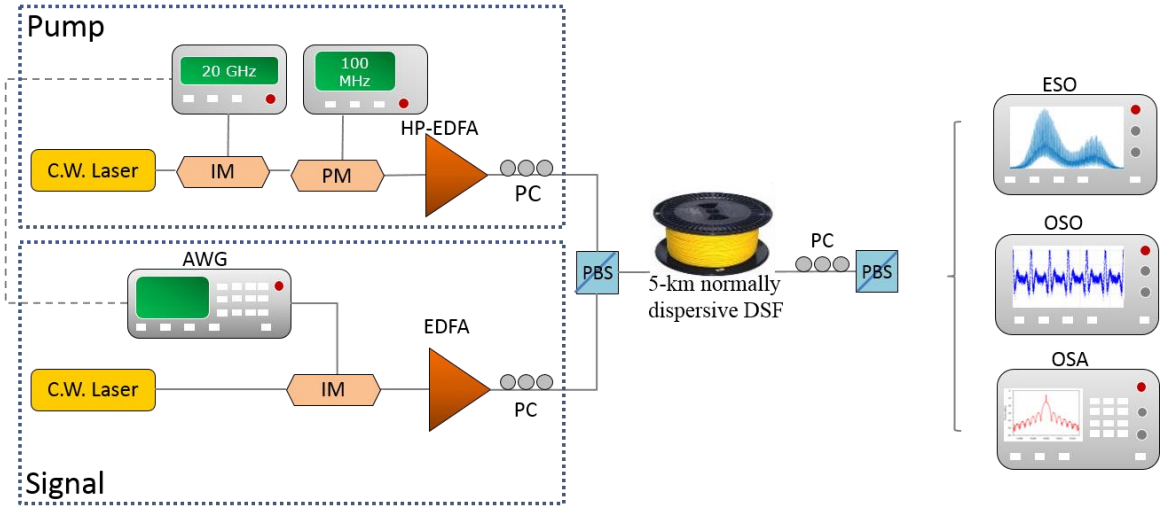

FIG. 2 Experimental setup. IM: intensity modulator, PM: phase modulator, AWG: arbitrary waveform generator, EDFA: Erbium doped fiber amplifier, PC: polarization controller, PBS: polarization beam splitter, ESO: electrical sampling oscilloscope, OSO: optical sampling oscilloscope and OSA: optical spectrum analyzer.

The experimental setup depicted in Fig. 2 was implemented in order to provide a proof-of-principle of this technique. For generating both the pump and signal waves, the scheme is initially divided into two arms. In the upper side (pump generation), an external cavity laser emitting at 1550 nm is first intensity modulated thanks to an intensity modulator (IM). A 20-GHz frequency clock is used to drive the IM at its null-transmission point so as to generate a pure carrier-suppressed 40-GHz sinusoidal wave. Subsequently, the resulting beating is phase modulated to 100 MHz thanks to a phase modulator (PM) so as to significantly increase the Brillouin threshold above the power levels involved in our experiment. On the other side (signal generation), another external cavity laser emitting at the same wavelength is intensity modulated by means of a second IM, which is driven by an arbitrary waveform generator (AWG). Therefore, the profile of this signal consists of a concatenation of two unbalanced Gaussian pulses with a duration of a few nanoseconds. Both the beat-signal (pump) and the arbitrary signal are independently amplified by two Erbium doped fiber amplifier (EDFA) and polarization multiplexed thanks to a polarization-beam splitter (PBS). The resulting signals are injected into a 5-km long optical fiber characterized by a chromatic dispersion $D = -2.5$ ps/nm/km at 1550 nm, an attenuation of 0.2 dB/km and a nonlinear Kerr coefficient $\gamma = 1.7$ $W^{-1}km^{-1}$. Finally, at the output of the system, the resulting pump and signal waves are polarization demultiplexed thanks to a second PBS and characterized in the time domain by means of two different oscilloscopes (an electrical sampling oscilloscope, ESO with a detection bandwidth of 70 GHz, and an optical sampling oscilloscope OSO from *Alnair Labs* having a temporal resolution of 1 ps). The output signal is also characterized in the spectral domain thanks to an optical spectrum analyzer (OSA).

The power of the incident signal is kept constant to 10 dBm while the pump power varied from 16 to 28 dBm. Figure 3a displays the normalized envelope of the arbitrary shaped nanosecond signal monitored at the output of the fiber when the sinusoidal pump wave is off (in red) and after the sampling process for an injected pump power of 26.5 dBm (in blue). We clearly observe a high-repetition-rate pulse train distributed along the signal wave and generated at the modulation frequency of the pump whose amplitude perfectly matches the initial signal envelope. We can thus notice that the sampling process acts in full strength whilst the shape of the signal is well preserved, in good agreement with our numerical predictions of Fig. 1c. It is noteworthy that for the results displayed in Fig. 3a, both signal and pump waves were synthetized thanks to the same initial RF clock and consequently appear temporally synchronized, which allows us to efficiently monitor the generation of the pulse train along the signal profile. Simultaneously, the sinusoidal pump wave has been broadened due to the combined effects of SPM and chromatic dispersion and undergoes an optical shock as illustrated in Fig. 3b for an injected power of 26.5 dBm, in full agreement with the trend reported in Fig. 1a.

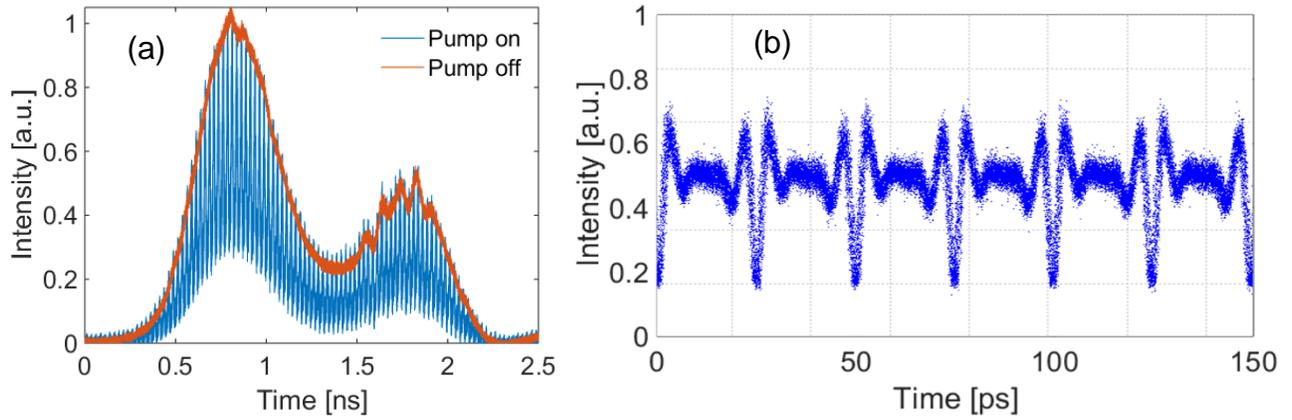

FIG. 3. (a) Normalized output signal in the temporal domain. For the red line, the pump is off and, for the blue curve, the signal is sampled (pump power equal to 26.5 dBm). This figure is obtained with the ESO. (b) Output intensity profile of the pump beam (power equal to 26.5 dBm). This snapshot is obtained with the OSO.

As it was previously mentioned, apart from the sampling process, the intensity profile of the signal wave is simultaneously magnified. The magnifying factor of the signal shape recorded at the output of the fiber as a function of the injected pump power has been reported in Fig. 4. Here, pump and signal waves are not temporally synchronized anymore. Therefore, the signal undergone an asynchronous sampling and only the envelope of the resulting magnified signal is then plotted. Fig. 4a compares the intensity profiles of a 1.5-ns arbitrary signal recorded at the output of the fiber with and without the presence of the 40-GHz sinusoidal pump wave, for different injected powers. We can notice that the output signal can reach a magnification factor close to 7 (8.5 dB) when the pump is activated. In particular, it is noteworthy that the second Gaussian pulse superimposed on the trailing edge of the arbitrary signal was hardly observable when the pump is off, whereas it can be easily detected when the signal is intensity magnified. In Figure 4b, we have reported the experimental magnifying factor (blue circles) of the output nanosecond signal as a function of the injected pump power. First of all, we can notice that the factor of magnification is not perfectly proportional to the pump power and that a maximum gain of 8.5 dB can be achieved owing to the present configuration. Experimentally, the pump power is here limited to 28 dBm because of the Brillouin backscattering effect. These experimental results have been compared both to numerical simulations (red solid line) and to our theoretical predictions (black crosses) assuming the simplified model described in the above section. We can observe an excellent agreement between our experimental measurements and both numerical as well as theoretical estimations, thus validating the present technique and its Manakov modeling.

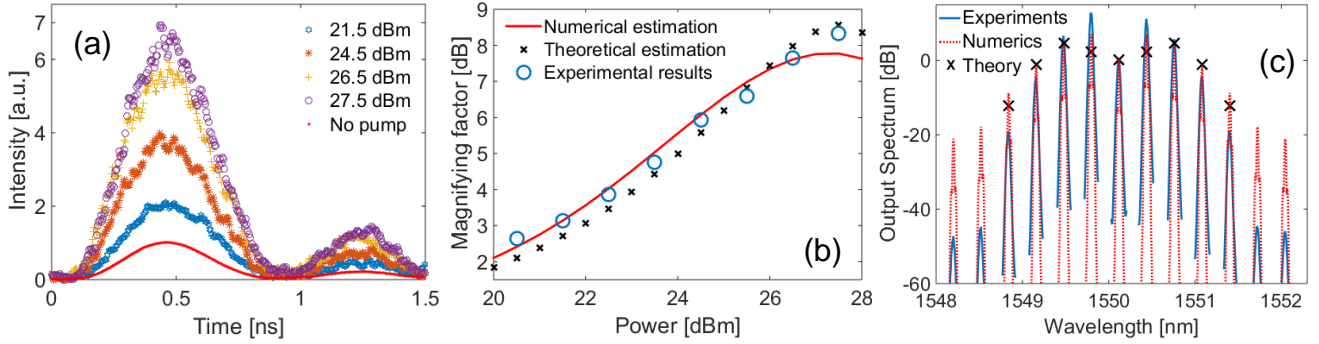

FIG. 4. (a) Output signal in the temporal domain. For the red line, the pump is off and the signal is magnified for different injected powers. This figure is obtained with the OSO and only the envelope is drawn. (b) Magnifying factor of the output signal as a function of the injected pump power. The experimental values (blue circles) are compared with the numerical simulations (red line) and theoretical predictions (black crosses). (c) Experimental spectra of the output signal (in blue) compared to theoretical (black crosses) and numerical estimations (red dashed line) for a pump power of 26.5 dBm.

Finally, the output signal is analyzed in the spectral domain. The phenomenon can be characterized as the generation of new frequency components at the pump frequency (40 GHz) related to the time-dependence of the nonlinear phase shift induced by XPM as illustrated in Fig. 4c. Moreover, we can observe once again a good agreement between the output recording spectrum and our theoretical predictions. Note that the disagreement among theoretical and numerical/experimental spectra grows larger as the pump power increases. This discrepancy at large pump power is related to the generation of new spectral components in the pump spectrum through FWM which, combined to chromatic dispersion, leads to severe distortion of its intensity profile; effect neglected in our present theoretical approach. Nevertheless, these components mainly affects the phase profile of the generated pulse signal, while not the magnitude. For this reason a good agreement between theoretical and numerical/experimental intensity profiles is still achieved for large pump powers (see Fig.1b).

In summary, we have proposed and experimentally demonstrated a novel technique enabling to simultaneously sample and magnify an arbitrary shaped optical signal. The present system is based on a periodic focusing effect upon an incident signal which takes place in a normally dispersive medium by means of a cross-phase modulation interaction induced by an orthogonally polarized high-repetition-rate sinusoidal pump wave. This localized periodic focusing induces a redistribution of energy along the signal profile which leads to the generation of a temporal comb at the repetition frequency of the pump whose amplitude is directly proportional to the incident signal, thus enabling its sampling and magnification. Using this technique, we have successfully implemented a 40-GHz sampling operation as well as an 8-dB magnification of an arbitrary shaped nanosecond signal around 1550 nm within a 5-km long normally dispersive optical fiber. Numerical and theoretical results are in good agreement with our experimental recordings. Alternatives based on wavelength multiplexed waves instead of polarization multiplexed waves could be developed, but the walk-off between the pump and signal may ultimately in such configuration degrade the efficiency of the magnification due to the focusing effect.

**Acknowledgements** This research is funded by the European Research Council under Grant Agreement 306633, ERC PETAL (www.facebook.com/petal.inside).